\documentstyle[preprint,revtex,eqsecnum]{aps}
\def\today{\ifcase\month\or
           January\or February\or March\or April\or May\or June\or
           July\or August\or September\or October\or November\or
           December\fi
           \space\number\day, \number\year}
\begin{document}
\draft
\preprint{U. of MD PP \#94-001}
\preprint{DOE/ER/40762-003}
\preprint{\today}
\begin{title}
Does $J/\psi \rightarrow \pi^{+} \pi^{-}$ fix the Electromagnetic Form
Factor\\ $F_{\pi}(t)$ at $t=M_{J/\psi}^2$?
\end{title}

\author{J. Milana}
\begin{instit}
Department of Physics, University of Maryland\\
College Park, Maryland 20742, USA\\
\end{instit}

\author{S. Nussinov}
\begin{instit}
Department of Physics, University of Maryland,
College Park, Maryland 20742\\
and Physics Department, Tel--Aviv University,
Ramat--Aviv, Tel--Aviv, ISRAEL\\
\end{instit}

\author{M. G. Olsson}
\begin{instit}
Physics Department, University of Wisconsin\\
Madison, Wisconsin 53706, USA\\
\end{instit}

\begin{abstract}
We show that the $J/\psi \rightarrow \pi^{+} \pi^{-}$ decay is a
reliable source of information for the electromagnetic form factor of
the pion at $t=M_{J/\psi}^2=9.6 {\rm GeV}^2$ by using general
arguments to estimate, or rather, put upper bounds on, the background
processes that could spoil this extraction.  We briefly comment on the
significance of the resulting $F_{\pi}(M_{J/\psi}^2)$.
\end{abstract}
\narrowtext

It is believed that the pion's electromagnetic form factor
$F_{\pi}(t)$ can be more reliablely calculated for $|t| \gg
\Lambda^2_{QCD}$ than the corresponding quantities for the nucleon.
However, $F_{\pi}(t)$ is more difficult to measure.\cite{undet} In
principle $F_{\pi}(t)$ can be measured for time like $t$ in $e^+e^-$
colliders via $e^+e^-
\rightarrow \pi^{+} \pi^{-}$.  However, for $t$ values of interest,
$|t| \approx 10$GeV$^2$, the above ratio is rather small and may be
difficult to extract from the few $e^+e^- \rightarrow \pi^{+} \pi^{-}$
events.\cite{epluseminusdata}

At the $J/\psi$ resonance the rate of all interactions is vastly
enhanced and branching ratios for rare channels such as the G--parity
(or isospin) forbidden $J/\psi \rightarrow \pi^{+} \pi^{-}$ can be
measured.  This rate could fix $F_{\pi}(t=M_{J/\psi}^2=9.6 {\rm
GeV}^2)$ if the decay proceeds predominantly via the one photon
exchange amplitude illustrated in Fig. 1a.  The dependence on the
charmonium's wavefunction can be eliminated by comparing the obtained
branching ratio $Br(J/\psi \rightarrow \pi^{+}\pi^{-})$ to the
leptonic decay rate $Br(J/\psi \rightarrow e^{+}e^{-})$, from which
one obtains that
\begin{equation}
\frac{Br(J/\psi \rightarrow \pi^{+}\pi^{-})}{Br(J/\psi
\rightarrow e^{+}e^{-})}=
\frac{F^{2}_{\pi}(M^2_{J/\psi})}{4}.
\end{equation}
The experimental values\cite{part.data.group}
\begin{eqnarray}
Br(J/\psi \rightarrow e^{+}e^{-}) &=& (6.27 \pm .20) 10^{-2}\\
Br(J/\psi \rightarrow \pi^{+}\pi^{-}) &=& (1.47 \pm .23) 10^{-4}
\label{pionrate}
\end{eqnarray}
would then imply that
\begin{equation}
F_{\pi}(M^2_{J/\psi}) = .098 \pm .008.\label{Fpi}
\end{equation}
This value of $F_{\pi}(t=9.6 {\rm GeV}^2)$ exceeds most theoretical
estimates\cite{asympt} \cite{C.Z.} and also the extrapolations of
$F_{\pi}(t)$ from the normally accepted values at large space--like
momentum inferred from $\pi$ electroproduction data (see however
\cite{undet}).\cite{pidata}

There are two additional mechanisms contributing to $J/\psi
\rightarrow
\pi^{+} \pi^{-}$:
\begin{equation}
A^{J/\psi \rightarrow \pi^{+} \pi^{-}} = A^{\pi}_\gamma +
A^{\pi}_{ggg} + A^{\pi}_{\gamma gg}.\label{amps}
\end{equation}
$A^{\pi}_{ggg}$ is taken to mean the contriubtuion to the amplitude of
a purely hadronic process, which perturbatively would be initiated via
a three gluon state and hence the nomenclature.  Likewise,
$A^{\pi}_{\gamma gg}$ is a mixed hadronic--electromagnetic
contribution that would be initiated via a two gluon, one photon
intermediate state (Figs. 1b, 1c respectively).  In the following we
will estimate $A^{\pi}_{ggg}$ and $A^{\pi}_{\gamma gg}$ and show that
both amplitudes fall considerablely short of explaining the observed
$J/\psi
\rightarrow \pi^{+} \pi^{-}$ decay rate, thus justifying
Eq.~(\ref{Fpi}) above.

\vglue 0.5cm

\underline{(I) $A^{\pi}_{ggg}$}:  Because the $J/\psi \rightarrow
\pi^{+} \pi^{-}$ violates isospin, this purely hadronic
process\cite{comment1} can proceed only via the isospin breaking
parameter $m^o_d - m^o_u$ which appears explicitly in the QCD
Lagrangian.\cite{vaffa} Such an amplitude should therefore be
suppressed by the small dimensionless factor $\epsilon_I = (m^o_d -
m^o_u)/Q$ with $Q$ some typical momentum in the problem.  Rather than
rely on any explicit, model dependent calculation, we present the
following more general argument by comparing with the SU(3) analog
process $J/\psi \rightarrow K \overline{K}$.  Since the $J/\psi
\rightarrow K \overline{K}$ decay violates SU(3)
symmetry,\cite{comment3} the corresponding purely hadronic decay
amplitude $A^{K}_{ggg}$ will have in this case the explicit small
SU(3) breaking suppression factor $\epsilon_{SU(3)} = (m^o_s -
m^o_{d,u})/Q$.  Consequently we expect that
\begin{equation}
\frac{A^{\pi}_{ggg}}{A^{K}_{ggg}} \approx
\frac{\epsilon_I}{\epsilon_{SU(3)}} = \frac{m^o_d - m^o_u}{m^o_s -
m^o_{d,u}} \approx .02-.03\label{hadratio}
\end{equation}
where in the spirit of the Vafa--Witten theorem\cite{vaffa} we used
the values of Lagrangian or ``current'' quark masses in estimating the
above ratio.  There are two $K \overline{K}$ decay modes, $J/\psi
\rightarrow K^o \overline{K^o}$ (or $K^o_s K^o_L$) and $J/\psi
\rightarrow K^+ K^-$.  The amplitude $A^{K}_{ggg}$ is simply given by
the former,
\begin{equation}
A^{K}_{ggg} \approx A^{J/\psi \rightarrow K^o_s
K^o_L}.\label{neutralkaons}
\end{equation}
The point is that the one photon and $\gamma gg$ contributions to the
$J/\psi \rightarrow K^o_s K^o_L$ decay also vanish in the SU(3) limit
due to the cancelling contribution of $s,\overline{d}$ quarks of
oppostite charge.\cite{kcharge} Thus the amplitudes $A^K_\gamma$ and
$A^K_{\gamma gg}$ are suppressed both by an explicit $\alpha_E$ and
$\epsilon_{SU(3)}$ factors and are hence negligible.  Multiplying
Eq.~(\ref{pionrate}) and Eq.~(\ref{hadratio}) with the observed
branching rate
\begin{equation}
Br(J/\psi \rightarrow K^o_s K^o_L) = (1.1 \pm .14) 10^{-4}
\end{equation}
implies that
\begin{equation}
A^{\pi}_{ggg} \approx \frac{1}{30}A^{J/\psi \rightarrow \pi^{+}
\pi^{-}}
\end{equation}
so that it can be safely ignored.

\vglue 0.5cm

\underline{(II) $A^{\pi}_{\gamma gg}$}:  It is very suggestive from a
perturbative framework that this process is suppressed by a factor of
$\alpha_s/\pi$ as it involves an extra gluon loop in comparison with
the corresponding expression for $A^{\pi}_{\gamma}$.  Indeed recent
detailed calculations\cite{Kahler} using a range of pion
wavefunctions\cite{asympt} \cite{C.Z.} indicate that
\begin{equation}
R = \frac{A^{\pi}_{\gamma gg}}{A^{\pi}_{\gamma}}
=\frac{\alpha_s}{\pi}\left\{
\begin{array}{c}.45\\.23\end{array}\right\} \approx
\left\{\begin{array}{c}\frac{1}{20}\\\frac{1}{40}\\\end{array}\right\}
\end{equation}
where the smaller $R$ value corresponds to the use of the more
realistic, non--asymptotic pion wave function\cite{C.Z.} allowing for
a larger $F_{\pi}(t)$ (which however still falls short by more than a
factor of two of explaining $Br(J/\psi \rightarrow \pi^{+}\pi^{-})$).

In order however not to rely too heavily on detailed model
calculations we would like to obtain a more general,
``phenomenological", estimate for $A^{\pi}_{\gamma gg}$.  Let us
therefore for the moment assume that only $A^{\pi}_{\gamma gg}$
contributes to the decay $J/\psi \rightarrow
\pi^{+} \pi^{-}$.

Consider first the total inclusive radiative decay of $J/\psi$ into
non--charmed hadrons: $Br(J/\psi \rightarrow \gamma + {hadrons})$.
This process can be viewed as $J/\psi \rightarrow \gamma gg$ with the
subsequent hadronization of the two gluon system, in the same way that
$J/\psi \rightarrow {hadrons}$ proceeds via a three gluon initial
perturbative state.  Thus the ratio
\begin{equation}
\frac{Br(J/\psi \rightarrow \gamma + { hadrons})} {Br(J/\psi
\rightarrow {hadrons \: only})} \approx \frac{Br(J/\psi \rightarrow
\gamma + gg)} {Br(J/\psi \rightarrow ggg)} =
\frac{16}{5}\frac{\alpha_E}{\alpha_s} = .07-.09,\label{photoratio}
\end{equation}
is readily\cite{ups} computed reflecting simply color and
symmetrization factors (and where we've taken $1/3\ge \alpha_s \ge
1/4$).  Note that the symmetrization factors {\it enhances} the case
with the final state photon by a factor of 3.  Such an enhancement
would in general be absent in the case that the bosons were not final
state particles but were instead found in a virtual intermediate
state, as we will be using below.  Nevertheless, in order to be as
conservative as possible, we will use Eq.~(\ref{photoratio}) in our
estimates without furthur modification.

For the three gluon system the incorporation of the gluons or the
quark pairs (to which they may convert) into hadrons is guaranteed by
the basic hypothesis of quark and gluon confinement.  However, we are
for our purposes interested in the case where the $\gamma gg$
intermediate state converts into hadrons {\it only}.  For this to
happen, the virtual photon must convert into a $q \overline{q}$ pair
which will cost an explicit extra factor of $\alpha_E$:\cite{comment2}
\begin{eqnarray}
Br(J/\psi \rightarrow \gamma gg \rightarrow hadrons) &=& Br(J/\psi
\rightarrow \gamma gg \rightarrow q\overline{q} gg \rightarrow
hadrons)\nonumber\\
&\approx& \alpha_E Br(J/\psi \rightarrow \gamma + hadrons) = (5 - 7)\:
10^{-4}.\label{qqgg}
\end{eqnarray}

We are focussing on a particular exclusive channel, namely a final
$\pi^+\pi^-$ state.  Thus we need to estimate the probability $f$ that
the $q\overline{q}gg$ state in Eq.~(\ref{qqgg}) hadronizes
specifically into a $\pi^+\pi^-$ state.  While it is uncertain how
reliablely one can directly compute $f$, we will infer an estimate for
$f$ from the probability that such a $q\overline{q}gg$ will hadronize
into an analogue $\pi \rho$ state, {\it i.e.} we will take that
\begin{equation}
f \equiv Br(q\overline{q} gg {\big|}_{J/\psi}  \rightarrow \pi^+\pi^-)
\approx \frac{1}{2}Br(q\overline{q} gg{\big|}_{J/\psi}  \rightarrow
\pi
\rho),\label{fcon}
\end{equation}
where the factor of $1/2$ reflects the two transverse polarizations of
the $\rho$ included in the $\pi \rho$ final state.

Note that the actual branching ratio
\begin{equation}
Br(J/\psi \rightarrow \pi^+ \rho^-) \approx 0.4 \%\label{rhopi}
\end{equation}
appears to be anomolously large in comparison with the branching ratio
to other two body channels.  Indeed it has triggered the speculation
of the existence of a glueball state in the vicinity of the
$J/\psi$.\cite{glueball} While such speculation is
controversial\cite{C.Z.}, there is general uniform agreement that the
$\pi^+ \rho^-$ branching ratio is unusually large.  Hence irrespective
of the correct explanation for $Br(J/\psi \rightarrow \pi^+ \rho^-)$,
its usage to estimate $A^{\pi}_{\gamma gg}$ must lead to a
conservative upper bound.  On the other hand, if the glueball
resonance scenario is correct, we would be severley overestimating
$A^{\pi}_{\gamma gg}$ since such a resonance would clearly not couple
to the $\gamma gg$ channel.

Finally, in order to estimate $A^{\pi}_{\gamma gg}$, we will
(conservatively) ignore the possible unusual behavior of the $\pi
\rho$ final state and note that a general two body, light meson
exclusive state is expected to be a short distance event.  Hence, in
order to generate the same $q\overline{q}gg$ state in Eq.~(\ref
{fcon}), we need to convert one gluon into a $q\overline{q}$ pair, and
thus we will take that
\begin{equation}
\frac{Br(J/\psi \rightarrow \gamma gg \rightarrow q\overline{q}
gg)}{Br(J/\psi \rightarrow ggg \rightarrow q\overline{q} gg)} =
\frac{1}{\alpha_s}Br(J/\psi \rightarrow \gamma gg \rightarrow
hadrons),\label{alphas}
\end{equation}
which will again enhance our estimate for $A^{\pi}_{\gamma gg}$ by
$1/\alpha_s$.  Combining Eqs.~(\ref{fcon}) and (\ref{alphas}) and
inserting (\ref{qqgg}) and (\ref{rhopi}), we obtain that
$A^{\pi}_{\gamma gg}$ alone would contribute a branching
\begin{equation}
Br^{\gamma gg}(J/\psi \rightarrow \pi^+ \pi^-) =  (3-6) 10^{-6},
\end{equation}
which is at least 25 times smaller than the observed value.  We hence
conclude that even under the most unfavorable scenarios,
$A^{\pi}_{\gamma gg}$ is less than a $20\%$ correction so that
\begin{equation}
A^{J/\psi \rightarrow \pi^{+} \pi^{-}} = A^{\pi}_\gamma +
A^{\pi}_{ggg} + A^{\pi}_{\gamma gg} \approx A^{\pi}_\gamma.
\end{equation}

Having established that the $J/\psi \rightarrow \pi^{+} \pi^{-}$ data
implies a fairly large value of $F_{\pi}(t=M_{J/\psi}^2)$, we briefly
turn to some concluding remarks:\\
\noindent (i) Recent results from E760 at Fermilab\cite{protondata}
indicates that the proton's electromagnetic form factor in the large
time--like region is also unusually large (by about a factor of $2$
over the space--like data).  A substantial imaginary part to hadronic
form factors in the time--like region could account for this
apparently systematic enhancement.\\
\noindent (ii) We expect that $A^{K}_\gamma \approx A^{\pi}_\gamma$ as
the kaon's and pion's electromagnetic form factors should be rather
similar at $t=M_{J/\psi}^2$.  Since $A^{K}_{\gamma gg}$ can be argued
to be small along similar lines presented for $A^{\pi}_{\gamma gg}$
and using our previous value for $A^{K}_{ggg}$,
Eq.~(\ref{neutralkaons}), we obtain that
\begin{equation}
A^{J/\psi \rightarrow K^+ K^-} \approx A^{J/\psi \rightarrow \pi^{+}
\pi^{-}} + A^{J/\psi \rightarrow K^o_s K^o_L}.\label{expectation}
\end{equation}
Considering that $A(J/\psi \rightarrow ggg \rightarrow K
\overline{K})$ is expected to have a substantial imaginary part (see
\cite{Kahler} for an explicit calculation of an analogous case), there
could in general be a large relative phase between the two terms in
Eq.~(\ref{expectation}).  Thus the latter is quite consistent with the
observed branching
\begin{equation}
Br(J/\psi \rightarrow K^+ K^-) = (2.4 \pm .3)10^{-4}.
\end{equation}

\noindent\underline{Acknowledgements}:  J.M. thanks S. J. Brodsky and
V. Chernyak for useful conversations.  S.N. thanks G. Karl for
discussions.  This work was supported in part by the U.S. Department
of Energy under grant No. DE-FG05-93ER-40762.

\figure{ The three contributions to the decay of charmonium into
$\pi^+
\pi^-$.  Curly lines are gluons, wavy lines photons.  (a) is
proportional to the pion's electromagnetic form factor.  (b) and (c)
are background processes not proportional to $F_\pi(M_{J\psi}^2)$.}

\begin{references}
\bibitem{undet}C. E. Carlson and J. Milana, Phys. Rev. Lett. {\bf 65},
1717 (1990).
\bibitem{epluseminusdata} L. M. Barkov, {\it et al.}, Nucl. Phys. {\bf
B256}, 365 (1984); DM2 Collaboration, Phys. Lett. {\bf 220B}, 321
(1989).
\bibitem{part.data.group}K. Hikasa {\it et al.}, Phys. Rev. D {\bf 45}
I.1 (1992) (Particle Data Group).
\bibitem{asympt}G. R. Farrar and D. R. Jackson, Phys. Rev. Lett {\bf
43}, 246 (1979); A. V. Efremov and A. V. Radyushkin, Phys. Lett. B
{\bf 94}, 245 (1980).
\bibitem{C.Z.}V. L. Chernyak and A. R. Zhitnitsky, Phys. Rep. {\bf
112}, 173 (1984).
\bibitem{pidata}C. J. Bebek {\it et al.}, Phys. Rev. D {\bf 17}, 1693
(1978).
\bibitem{comment1} Indeed our arguments for estimating $A^{\pi}_{ggg}$
will not rely on the specific 3 gluon mechanism, but only on the fact
that this is a purely hadronic amplitude.
\bibitem{vaffa} The Vafa--Witten theorem [C. Vafa, and E. Witten,
Nucl.  Phys. {\bf B234}, 173 (1984)] based on QCD inequalities exclude
a spontaneous breaking of isospin or other global vectorial
symmetries.  Any isospin violating purely hadronic amplitude must
therefore contain an explicit factor of $m^o_d - m^o_u$.
\bibitem{comment3} Recall that in the SU(3) flavor symmetry limit,
$A(J/\psi \rightarrow K^+ K^-) = A(J/\psi \rightarrow \pi^{+}
\pi^{-})$ .
\bibitem{kcharge} Indeed all electromagnetic properties such as the
charge radius of the $K^o$ vanish in the SU(3) limit.  See e.g., O. W.
Greenberg, S. Nussinov, and J. Sucher, Phys. Lett. {\bf 70B}, 289
(1977).
\bibitem{Kahler} R. Kahler and J. Milana, Phys. Rev. D {\bf 47}, R3690
(1993).
\bibitem{ups}P. B. Mackenzie and G. P. Lepage, Phys. Rev. Lett. {\bf
47}, 1244 (1981).
\bibitem{comment2} The modification due to the fact that empirically
$Br(J/\psi \rightarrow {hadrons \: only}) \approx 2/3$ will cancel in
our estimates between Eqs.~(\ref{qqgg}) and (\ref{rhopi}) below.
\bibitem{glueball} Wei--Shou Hou and A. Soni, Phys. Rev. Lett. {\bf
50}, 569 (1983); S. J. Brodsky, G. P. Lepage and San Fu Tuan, Phys.
Rev. Lett. {\bf 59}, 621 (1987).
\bibitem{protondata} T. A. Armstrong {\it et al.}, Phys. Rev. Lett.
{\bf 70}, 1212 (1993).
\end{references}
\end{document}